\documentstyle[12pt,aasms4]{article}
\begin{document}

\title{The Effects of Thermonuclear Reaction Rate Variations on Nova Nucleosynthesis: A Sensitivity Study}

\author{Christian Iliadis}
\affil{Department of Physics and Astronomy, The University of North Carolina at Chapel Hill, Chapel Hill,
North Carolina 27599--3255; and Triangle Universities Nuclear Laboratory, 
Durham, North Carolina 27708--0308; iliadis@unc.edu}

\author{Art Champagne}
\affil{Department of Physics and Astronomy, The University of North Carolina at Chapel Hill, Chapel Hill,
North Carolina 27599--3255; and Triangle Universities Nuclear Laboratory, 
Durham, North Carolina 27708--0308; aec@tunl.duke.edu}

\author{Jordi Jos\'e}
\affil{Departament de F\'{\i}sica i Enginyeria Nuclear (UPC), Avinguda
V\'{\i}ctor Balaguer, s/n, E-08800 Vilanova i la Geltr\'u (Barcelona),
SPAIN; and
Institut d'Estudis Espacials de Catalunya,
Edifici Nexus-201, C/Gran Capit\`a 2-4, E-08034 Barcelona, SPAIN; jjose@ieec.fcr.es}

\author{Sumner Starrfield}
\affil{Department of Physics and Astronomy, Arizona State University, Tempe, Arizona 85287--1504; 
sumner.starrfield@asu.edu}

\and

\author{Paul Tupper}
\affil{Scientific Computing -- Computational Mathematics Program,
Stanford University, Stanford, California
94305; tupper@sccm.stanford.edu}

\begin{abstract}

We investigate
the effects of thermonuclear reaction rate uncertainties on nova nucleosynthesis. 
One--zone nucleosynthesis calculations have been performed by adopting
temperature--density--time profiles of the hottest hydrogen--burning zone 
(i.e., the region in which most of the nucleosynthesis takes place). We obtain our profiles 
from 7 different, recently published, hydrodynamic nova simulations covering peak
temperatures in the range from T$_{peak}$=0.145--0.418 GK. For each of these profiles, we individually 
varied the rates of 175 reactions within their associated errors and 
analyzed the resulting abundance changes of 142 isotopes in the mass range below A=40.
In total, we performed $\approx$7350 nuclear reaction network calculations.
We use the most recent thermonuclear reaction rate
evaluations for the mass ranges A=1--20 and A=20--40. For the theoretical astrophysicist, our results 
indicate the extent to which nova nucleosynthesis calculations depend on presently uncertain
nuclear physics input, while for the experimental nuclear physicist our results represent at least a 
qualitative guide for future measurements at stable and radioactive ion beam facilities.
We find that present reaction rate estimates are reliable for predictions of
Li, Be, C and N abundances in nova nucleosynthesis. However, rate uncertainties of several reactions
have to be reduced significantly in order to predict more reliable O, F, Ne, Na, Mg, Al, Si, S,
Cl and Ar abundances. Results are presented in tabular form for each adopted nova simulation.
\end{abstract}

\keywords{nuclear reactions, nucleosynthesis, abundances -- stars: novae}

\section{INTRODUCTION}

Classical novae occur in binary star systems consisting of a white dwarf and
a main sequence star. When the companion star fills its Roche lobe,
matter passes through the inner Lagrangian point and accumulates in an accretion disk
before falling onto the white dwarf. The accreted layer gradually grows in mass.
For sufficiently 
small mass--accretion rates, the deepest layers of the accreted material become partially 
degenerate. The temperature in the accumulated envelope increases because of 
compressional heating and energy release from nuclear reactions until a thermonuclear 
runaway occurs. 
At some time during the evolution, material from 
the white dwarf core is mixed into the accreted hydrogen--rich layer.
As a consequence, a significant fraction of material, enriched in the products 
of hot hydrogen burning, is ejected into the interstellar medium. Spectroscopic studies of
classical novae show enrichments of either C, N, O or 
of certain elements in the range from Ne to Ar
(Gehrz et al. 1998, and references therein; Starrfield et al. 1998).
The observed abundance patterns have been explained by assuming that the outbursts involve two 
fundamentally different types of white dwarfs with a composition consisting primarily of either 
carbon and oxygen (CO) or oxygen and neon (ONe).

The study of classical novae is of considerable interest for several reasons. 
First, spectroscopic studies of nova ejecta, when properly interpreted, reveal the composition of the
underlying white dwarf, thereby constraining models of stellar evolution. 
Second, the observed
elemental abundances also reflect the evolution of the thermonuclear runaway, such as peak 
temperatures and expansion time scales, and thus provide constraints for models of
stellar explosions (Starrfield et al. 1998, 2000). 
Third, classical novae clearly contribute to the chemical evolution of the Galaxy. In fact, they 
have been proposed as the major source of the isotopes $^{13}$C and $^{17}$O, and perhaps $^{15}$N 
(Jos\'{e} \& Hernanz 1998). They may also represent a site
for production of the cosmologically interesting isotope $^{7}$Li (Arnould \& Norgaard 1975;
Starrfield et al. 1978; Hernanz, Jos\'{e}, Coc, \& Isern 1996) as suggested by recent models of Galactic
chemical evolution (Romano et al. 1999).
Fourth, it is believed that radioactive isotopes are synthesized in nova outbursts. 
Short--lived isotopes, such as $^{14}$O ($\tau$$_{1/2}$=71 s), $^{15}$O ($\tau$$_{1/2}$=2 min) and 
$^{17}$F ($\tau$$_{1/2}$=65 s), can reach the outer layers of the 
accreted envelope via convection, and their $\beta$--decays provide an important energy source
for the ejection of material (Starrfield et al. 1972).
The decays of
the short--lived nuclei $^{13}$N ($\tau$$_{1/2}$=10 min) and $^{18}$F ($\tau$$_{1/2}$=110 min) produce 
$\gamma$--radiation of 511 keV and below, related to electron--positron annihilation and 
Compton--scattering, 
at a time when the expanding envelope becomes transparent to 
$\gamma$--rays (G\'{o}mez--Gomar et al. 1998; Hernanz et al. 1999). The decays of the longer--lived 
isotopes $^{7}$Be ($\tau$$_{1/2}$=53 d) 
and $^{22}$Na ($\tau$$_{1/2}$=2.6 y) produce $\gamma$--rays with energies of E$_{\gamma}$=478 and 1275 keV,
respectively (Clayton \& Hoyle 1974; Leising \& Clayton 1987). Observations of $\gamma$--rays from novae 
have been attempted 
with several satellites, but no positive detection has been reported. In the 
near future, however, novae will be promising targets for more sensitive instruments, such as the 
International Gamma--Ray Astrophysics Laboratory (INTEGRAL).
Fifth, the discovery of $^{26}$Al ($\tau$$_{1/2}$=7.4$\times$10$^{5}$ y) in the interstellar medium
(Mahoney et al. 1982) provided direct proof that nucleosynthesis is currently active in the Galaxy. 
>From the observed intensity of the 1809 keV $\gamma$--ray line emission, it has been estimated that the 
production rate of $^{26}$Al in the Galaxy is $\approx$2 M$_{\odot}$ per 10$^{6}$ y. 
Although massive stars have been proposed as the main source of $^{26}$Al (Diehl et al. 1995; Prantzos \& Diehl 1996;
Diehl 1997; Kn\"{o}dlseder 1999), 
a contribution from classical novae cannot be ruled out (Politano et al. 1995;
Jos\'{e}, Hernanz, \& Coc 1997).
Sixth, the recent discovery of several presolar SiC grains with anomalous C, N, Al and Si isotopic ratios
points towards a nova origin (Amari et al. 2001). If this identification is accurate, then the measured 
isotopic composition provides important constraints on both the nucleosynthesis and on the conditions in 
stellar outflows and circumstellar grain formation (Gehrz et al. 1998).

The thermonuclear runaway model reproduces several key features observed in nova outbursts. 
At present, the most successful calculations involve one--dimensional hydrodynamic codes that
are directly coupled to large nuclear reaction networks (Kovetz \& Prialnik 1997; Jos\'{e} \& Hernanz 1998; 
Starrfield et al. 1998, 2000; and references therein). 
However, some outstanding problems remain to be solved
(Gehrz et al. 1998; Jos\'{e} \& Hernanz 1998; Starrfield et al. 1998, 2000). For
example, the masses of the underlying white dwarfs are unknown and the rates of mass accretion are poorly
constrained. The composition of white dwarfs involved in either CO or ONe novae is far from understood
and may vary from outburst to outburst. 
The mechanism responsible for the mixing of white dwarf core material into the
accreted hydrogen envelope is not universally accepted. The amount of mass ejected is controversial. 
Finally, many nuclear reaction cross sections
entering in the hydrodynamic model calculations are uncertain by orders of magnitude. 

In the present work, we focus on the effects of reaction rate uncertainties in nova model calculations. 
In the past, such effects were frequently ignored by stellar modelers who used only one specific 
set of recommended reaction rates from available libraries. Reaction rate uncertainties 
in hydrodynamic nova model calculations have been rarely explored in previous work. These studies
were mainly concerned with the effects of a few uncertain reaction rates on the production of specific
isotopes of interest, such as $^{18}$F (Coc et al. 2000), 
$^{22}$Na and $^{26}$Al (Jos\'{e}, Coc, \& Hernanz 1999), and Si--Ca 
(Jos\'{e}, Coc, \& Hernanz 2001). In the present work, we 
describe a more extensive approach. We independently vary the rates of 175 reactions 
that participate in nova model nucleosynthesis and analyze the resulting abundance variations of 
142 isotopes in the mass range below A=40. In our calculations we take advantage of the two most recent 
thermonuclear reaction rate evaluations for the mass ranges A=1--20 (Angulo et al. 1999) and A=20--40 
(Iliadis et al. 2001). 
For the theoretical astrophysicist, our results indicate the extent to which the nucleosynthesis depends on 
presently uncertain nuclear physics input, while for the experimental nuclear physicist our results represent
at the least a qualitative guide for future measurements at stable or radioactive ion beam facilities.

Our philosophy and general issues related to the present work are described in $\S$ 2. 
In $\S$ 3 we explain our strategy and procedures in more detail. Results are presented in $\S$ 4 and
discussed in $\S$ 5. A summary and conclusions are given in $\S$ 6.

\section{PHILOSOPHY}

Experimental nuclear physicists frequently inquire about ``the most important nuclear reaction
to be measured in order to explain the nova phenomenon". Recalling the discussion in the last section, 
we can state with confidence that a single most important reaction does not exist. Rather, 
{\it different key reactions are important for different aspects of nova nucleosynthesis}.
We also need to make an important distinction. Of principle interest is not the identification 
of the most important nuclear reactions (for example, those which produce most of the energy), but the 
search for
those reaction rate {\it uncertainties} which have the largest impact on nova simulations. It is precisely
these reaction rate uncertainties which need to be addressed with significant
new measurements.

Modern reaction networks used in nova studies typically involve $\approx$100 isotopes linked by
$\approx$1000 nuclear reactions and decays. The situation is very complex and intuitive guesses
regarding the most important reaction rate uncertainties are inadequate. 
Clearly, a quantitative approach is needed.
Consider as an example a nuclear reaction with a rate uncertainty of a factor of 100 in
the temperature range of interest. The most direct approach to investigate the effects
of this uncertainty on the overall nucleosynthesis would require several hydrodynamic
simulations. The first calculation might be performed with the recommended
rate for this particular reaction, while in subsequent calculations the rate might be changed 
by specific factors within the quoted uncertainty. A decision regarding the ``importance"
of the reaction rate error under consideration can then be based, for example, on the extent of
isotopic abundance variations predicted by these calculations.
This procedure would then be repeated for all other nuclear reactions which are of potential
interest for nova nucleosynthesis. At this point, it has to be kept in mind that a single
hydrodynamic nova calculation typically takes several CPU hours on present--day computers.
Although the approach described above is useful for a relatively small number of reaction rate
changes (Jos\'{e}, Coc, \& Hernanz 1999; Coc et al. 2000; Jos\'{e}, Coc, \& Hernanz 2001), 
it is clear that it
is not suitable for purposes of the present work because of limitations in computing time.

In the present work, a different approach is utilized. Our calculations are performed
with an extended reaction network by using temperature--density--time profiles extracted
from recent hydrodynamic nova simulations. The advantage of this procedure is that a single network
calculation lasts only a few minutes. This has allowed us to independently vary the rates of 175 reactions
by different factors within their uncertainties, and to analyze the resulting abundance variations of 
142 isotopes in the mass range below A=40. The procedure is repeated for a number of
temperature--density--time profiles obtained from recent hydrodynamic nova simulations involving different
white dwarf masses and compositions.
In total, we have performed $\approx$7350 reaction network calculations.
We would also like to point out a disadvantage of this procedure. The reaction network
is not coupled directly to the hydrodynamics and, consequently, we ignore the important 
effect of convection on the final nova abundances. As pointed out previously (see, for example, 
Lazareff et al. 1979 or Jos\'{e} \& Hernanz 1998), convective mixing 
carries material from the hydrogen burning region to the surface on short time scales. This will cause an 
increase in ejected abundances of fragile nuclei that would have been destroyed, if they had not been carried 
to higher and cooler layers. Therefore, our calculations are unsuitable for defining {\it absolute}  
isotopic abundances resulting from nova nucleosynthesis. However, we claim that our procedure is 
adequate for exploring the effects of reaction rate uncertainties on abundance {\it changes} in the hottest 
hydrogen burning zone, i.e., the region in which most of the nucleosynthesis takes place.
For a few selected cases, we have compared the results of our one--zone (or ``co--processing") 
nucleosynthesis calculations with those obtained by the hydrodynamic code coupled directly to the 
reaction network. As will be seen, the results are in reasonable agreement.
Similar approaches\footnote{A recent article by Hix et al. (2002) also addresses effects of reaction rate 
uncertainties in nova nucleosynthesis. They assign random errors to each reaction rate in their network 
by using Monte Carlo
techniques. Their procedure represents a complementary approach to a similar problem.} 
investigating the nucleosynthesis in solar models (Bahcall et al. 1982) and in massive stars 
(The et al. 1998; Hoffman, Woosley, \& Weaver 2001) have been reported
previously. 

Finally, we would like to address an issue which some of us have confronted in the past.
One might argue that it is of little use to identify key reaction rate errors since
hydrodynamic nova modeling carries significant uncertainties ($\S$ 1).
However, it must be emphasized that the abundances observed in nova ejecta or in
presolar grains from novae
provide strong constraints for nova simulations because nuclear reactions are very
sensitive to temperature. Clearly, such constraints are only
useful for improving current stellar models if key nuclear reaction rates are known with sufficient
accuracy.

\section{STRATEGY}

\subsection{Nuclear Reaction Network}

The nuclear reaction network used in the present work follows the detailed evolution of 142 stable 
and proton--rich isotopes from hydrogen to calcium. 
For the physical conditions achieved by the nova models adopted in our work, this network is
appropriate for nucleosynthesis calculations. The assumption is supported by the fact that
overabundances of elements beyond calcium are not observed in nova ejecta.
The nuclei are linked by 1265 nuclear processes
including weak interactions, reactions of type (p,$\gamma$), (p,$\alpha$), ($\alpha$,$\gamma$) 
etc., and corresponding reverse reactions. 

For the construction of the thermonuclear reaction rate library we have used, with few exceptions, 
the most recent compiled and evaluated results given in Angulo et al. (1999) and Iliadis et al. (2001) 
for the mass ranges A=1--20 and A=20--40, respectively. 
For the reactions $^{8}$B(p,$\gamma$)$^{9}$C\footnote{The astrophysical S--factor for this reaction  
has been estimated recently by measuring the proton--transfer $^{8}$B(d,n)$^{9}$C (Beaumel et al. 2001).
The new reaction rate estimate is smaller by a factor of 4 compared to the results of Wiescher et al.
(1989) which are used in the present work. This difference is unimportant for nova nucleosynthesis 
(see $\S$ 5).
}, $^{9}$C($\alpha$,p)$^{12}$N, $^{8}$B($\alpha$,p)$^{11}$C,
$^{11}$C(p,$\gamma$)$^{12}$N and $^{12}$N(p,$\gamma$)$^{13}$O we used the reaction rates of
Wiescher et al. (1989). For the reaction $^{17}$F(p,$\gamma$)$^{18}$Ne
we adopted the results of Bardayan et al. (2000), while for $^{17}$O(p,$\gamma$)$^{18}$F and 
$^{17}$O(p,$\alpha$)$^{14}$N we made use of the rates from Blackmon et al. (2001). The reaction
rates for $^{18}$F(p,$\gamma$)$^{19}$Ne and $^{18}$F(p,$\alpha$)$^{15}$O were taken from Coc et al. 
(2000). For the reactions $^{13}$C(p,$\gamma$)$^{14}$N,
$^{14}$N(p,$\gamma$)$^{15}$O, $^{16}$O(p,$\gamma$)$^{17}$F, $^{18}$O(p,$\alpha$)$^{15}$N,
$^{19}$F(p,$\gamma$)$^{20}$Ne, $^{19}$Ne(p,$\gamma$)$^{20}$Na, $^{15}$O($\alpha$,$\gamma$)$^{19}$Ne and
$^{14}$O($\alpha$,p)$^{17}$F we still employ the rates from Caughlan \& Fowler (1988) since changes
in recent updates are small (less than 30\%). Our network also includes all $\beta$--delayed\footnote{
Consider, for example, the positron decay of $^{29}$S. Previous network calculations included only
the link $^{29}$S$\rightarrow$$^{29}$P which represents the $\beta$--decay to the $^{29}$P
ground state. However, the nucleus $^{29}$S also $\beta$--decays with about equal probability
to excited $^{29}$P states which are unbound. These levels decay subsequently via proton emission,
leading to the final nucleus $^{28}$Si. Clearly, the $\beta$--decay $^{29}$S$\rightarrow$$^{29}$P
and the $\beta$--delayed proton decay $^{29}$S$\rightarrow$$^{28}$Si compete with each other and have 
to be treated as separate links in the network.
} 
proton and $\alpha$--particle decays in the mass range of interest. All partial half--lives for 
$\beta$--decays and $\beta$--delayed decays have been adopted from the recent NUBASE evaluation
(Audi et al. 1997). The ground and isomeric state of $^{26}$Al
have been treated as separate nuclei (Ward \& Fowler 1980) 
and the communication between those states through thermal
excitations involving higher--lying excited $^{26}$Al levels has been taken into account explicitly.
The required $\gamma$--ray transition probabilities have been adopted from Runkle et al. (2001).
The library used here for nucleosynthesis calculations for the mass range A$\leq$40 is, in our opinion, 
the most recent and 
consistent set of thermonuclear reaction rates available at present.

\subsection{Temperature--Density Evolution and Initial Composition}

In addition to the information described above, our one--zone reaction network calculations require
assumptions regarding the evolution of temperature and density, and the initial envelope composition. 

In the present work, we have used temperature--density--time profiles of the hottest hydrogen burning zone, 
obtained from recently published hydrodynamic nova simulations. 
Properties of these evolutionary nova models are summarized in Table 1 and
are described in detail elsewhere (Politano et al. 1995; Jos\'{e}, Coc, \& Hernanz 1999; Starrfield 
et al. 2002).
Stellar evolution theory predicts that the masses of CO (ONe) white dwarfs are smaller (larger)
than $\approx$1.1M$_{\odot}$. Therefore, we have considered several models of CO and ONe novae with 
white dwarf masses of 0.8--1.0M$_{\odot}$ and 1.15--1.35M$_{\odot}$, respectively. The corresponding 
temperature--density--time profiles are displayed in Figure 1. Note, that ONe nova model S1 of Starrfield 
et al. (2002) was calculated with the same thermonuclear reaction rate library as used in the present work, 
while the nova models of Politano et al. (1995) and Jos\'{e}, Coc, \& Hernanz (1999) were calculated 
with previous reaction rate libraries.

Our network calculations, for a specific temperature--density--time profile, have been performed with
the same initial isotopic composition as was used in the corresponding hydrodynamic nova
simulation (Table 1). Initial isotopic abundances (in mass fractions) 
for the nova models considered here are listed in 
Table 2 and are also displayed in Figure 2. We would like to point out that the initial abundances 
employed in the ONe nova models of Politano et al. (1995) and Starrfield et al. (2002) differ 
significantly from those of Jos\'{e}, Coc, \& Hernanz (1999). Therefore, we are also studying the effects of
different initial compositions on the final abundance changes.

\subsection{Reaction Rate Errors and Reaction Rate Variations}

The investigation of reaction rate sensitivities in nova nucleosynthesis requires the variation
of reaction rates within their respective uncertainties. Therefore, quantitative estimates of
reaction rate errors are needed. For a subset of reactions considered here, we list in Table 3 
reaction rate errors adopted in the present work.
For most reaction rates involving stable or long--lived target nuclei the errors were taken from either
Angulo et al. (1999) or from Iliadis et al. (2001). For $^{17}$O+p we use the errors of Blackmon et
al. (2001), since new experimental results have become available. The reader should realize that 
it is frequently difficult to assign errors to reaction rates. This situation arises, for example,
if Hauser--Feshbach theory is used to calculate a reaction rate, or if a reaction involves a 
short--lived 
target nucleus. In the former case\footnote{The reactions of interest
here involve light target nuclei (A$\leq$40) and have small Q--values (Q$\leq$10 MeV).
Therefore, we expect Hauser--Feshbach reaction rates to provide results in excess of the
usually quoted ``factor of 2 reliability" (Hoffman et al. 1999; Rauscher
\& Thielemann 2000). This point has been discussed in more detail by Iliadis et al. (2001).},
we have generally assumed that reaction rate errors amount to 
a factor of 100 up and down. The same assumption has been made in the latter case as well,
with a few important exceptions. The reaction $^{8}$B(p,$\gamma$)$^{9}$C has not been measured
directly, but the corresponding reaction rates can be estimated by using results of a recent
proton--transfer reaction study (Beaumel et al. 2001). In this case we assumed a conservative
reaction rate error of a factor of 10.
For the $^{13}$N(p,$\gamma$)$^{14}$O reaction rates we adopted the errors of
Angulo et al. (1999), while for $^{18}$F(p,$\gamma$)$^{19}$Ne and 
$^{18}$F(p,$\alpha$)$^{15}$O we used the errors of Coc et al. (2000). Bardayan et al. (2000) report an 
error of only 15\% for the $^{17}$F(p,$\gamma$)$^{18}$Ne reaction rates at nova temperatures. However, 
it must be
emphasized that the proton capture reaction on $^{17}$F has not been measured. In our opinion, 
an error of a factor of 10 is a more realistic estimate. Finally, Iliadis et al. (1999) report 
reaction rate errors of a factor of 2 for the proton captures on $^{27}$Si, $^{31}$S, $^{35}$Ar and $^{39}$Ca. 
In the present work, we adopted a more conservative error of a factor 10.
In some cases, reaction rate uncertainties 
are not constant but depend on stellar temperature (for example, see Figs. 2--4 in Iliadis et
al. 2001). If a reaction rate error varied significantly with temperature, for the sake of simplicity 
we have adopted in our 
network calculations the maximum reaction rate error in the
temperature range of interest to nova nucleosynthesis (T=0.1--0.4 GK). This assumption is conservative
since it can overestimate some of our predicted abundance changes.

Among the 1265 nuclear processes included in our network, we varied the rates of 175 selected reactions.
Those included all exothermic (p,$\gamma$) and (p,$\alpha$) reactions and the most important 
($\alpha$,$\gamma$) and ($\alpha$,p) reactions, on stable and proton--rich target nuclei with masses 
A$\leq$40. Only a subset of 62 reactions is listed in Table 3. The rates of those 175 reactions, together 
with the corresponding reverse reaction rates, have been varied individually by factors of 
100, 10, 2, 0.5, 0.1 and 0.01 in successive reaction network calculations. Since we have explored nova 
nucleosynthesis for seven different temperature--density--time profiles (Table 1), a total of 
175$\times$6$\times$7=7350 network calculations were performed.

\section{RESULTS}

For each network calculation, the final abundances of 142 isotopes were
analyzed. Short--lived isotopes (e.g., $^{13}$N, $^{14}$O, $^{15}$O and $^{17}$F) 
present at the end of a network calculation were assumed to decay to their stable daughter
nuclei. 

In Table 4 we list the final isotopic abundances (in mass fractions) for each temperature--density--time profile
considered in the present work (Table 1). These results have been obtained by using recommended rates
for all reactions in our network, as discussed in $\S\S$ 3.1.
We emphasize again that, for reasons given in $\S$ 2, the abundances presented in Table 4 should 
not be directly compared to abundances observed in nova ejecta or to those obtained from a full
hydrodynamic calculation. Table 4 is mainly useful for the purpose of comparing
final abundances from different one--zone nucleosynthesis calculations.

The results of our reaction rate variation procedure are presented in Tables 5--11.  For each 
temperature--density--time profile we list 
in column (1) the reaction whose rate has been varied, in column (2) the isotope i whose abundance
changed because of the rate variation, and in columns (3)--(8) the factor change, X$_{i}$/X$_{i,rec}$, 
in final isotopic abundance for rate variations by factors of 100, 10, 2, 0.5, 0.1 and 0.01.
Specifically, X$_{i,rec}$ refers to the final isotopic abundance
of isotope i obtained from a network calculation involving recommended rates only; 
X$_{i}$ refers to the final isotopic abundance of isotope i obtained from a network calculation
in which the rate of a single reaction (listed in column 1) has been multiplied by a specific factor. 
Only significant final abundance changes are presented.
Results have been listed only if i) a final abundance changed by at least 
10\% compared to the reference calculation performed with our recommended reaction rate library 
($\S\S$ 3.1), and ii) the reaction rate was varied by a factor less than (or close to) the assigned 
reaction rate error (Table 3). In Figure 3 we display the results of reaction rate variations
for a few selected cases only. Our results are discussed in $\S$ 5.

\section{DISCUSSION}

We start the discussion with two necessary (but not sufficient) conditions that have to be fullfilled 
for the experimental nuclear physicist in order to perform a meaningful new measurement of a 
particular nuclear reaction. First, the nuclear reaction must have a {\it significant influence on a stellar 
model property that can be related to an astronomical observable}. Second, the nuclear reaction rate must 
have an {\it error giving rise to a significant uncertainty of a stellar model property}. 
The observable could be an isotopic abundance, a luminosity or a mass ejection velocity. In this section 
we will not attempt to discuss all of the results listed in Tables 5--11, but concentrate on those cases for which 
the two conditions outlined above apply.

As a first example, we consider the $^{39}$K(p,$\gamma$)$^{40}$Ca reaction. According to Table 3, we assign a 
factor of 100 uncertainty to the reaction rate. Increasing the recommended rate for this 
reaction by a factor of 100 decreases the final $^{39}$K abundance in all of our ONe nova network 
calculations by more
than an order of magnitude. However, potassium has not yet been observed in nova ejecta. In this case, the 
first condition is not fullfilled and, therefore, calculated potassium abundances are unimportant 
for testing 
current nova models. We will not discuss such cases further. Nevertheless, the results are listed
in Tables 5--11 since future observations of nova ejecta could perhaps reveal the presence of elements like 
potassium.

As another example, consider the $^{20}$Ne(p,$\gamma$)$^{21}$Na reaction. The error for this rate is 
about 70\% (Table 3). Rate variations by factors of 2 and 0.5 produce final abundance changes 
for any isotope of less than a factor of 2. These calculated abundance changes are close 
to present uncertainties of observed abundances
in nova ejecta. In this case, the second condition is not fullfilled and, consequently, 
it is unlikely that a new and improved measurement of this reaction will provide additional constraints for 
current nova models. Again, we do not discuss such cases further but list the results in Tables 5--11
since abundances observed in nova ejecta are likely to become more precise in the future.

It is important to point out that the reaction--rate variations performed in the present work
have only a minor influence on the amount of hydrogen consumed, the amount of helium produced, and
the total thermonuclear energy released.
In the following, we will focus on final isotopic abundance changes of those elements that are considered 
the most important for nova nucleosynthesis ($^{7}$Li, $^{7}$Be, C, N, O, $^{18}$F, Ne, Na, Mg, Al, Si, 
S, Cl and Ar; 
see $\S$ 1). The variation of reaction rates within their assigned error in the temperature range T=0.1--0.4 GK
($\S\S$ 3.3 and Table 3) 
will be referred to simply as ``reaction rate variations". When using the expression ``abundance" we 
mean more specifically the final isotopic abundance obtained at the end of a network calculation.
Furthermore, we have regarded abundance changes as significant only if they
amount to at least a factor of 2.
The mass regions A$<$20 and A$\geq$20 are discussed separately in the next subsections.

\subsection{Mass Region A$<$20}

\subsubsection{Isotopes $^{7}$Li and $^{7}$Be}

In explosive hydrogen burning, the isotope $^{7}$Li is produced by the decay of $^{7}$Be.

The isotopic abundance of $^{7}$Be depends only weakly on reaction rate variations
in CO nova models. Abundance changes amount to less than a factor of 2 and, therefore,
cannot be regarded as significant.

In ONe nova models P1 and P2, the abundance of $^{7}$Be is also insensitive to reaction rate
variations. For models JCH1 and JCH2, the $^{7}$Be abundance changes by less than a factor of
2 as a result of varying the $^{7}$Be(p,$\gamma$) reaction rates within adopted errors\footnote{
Abundance changes of $^{7}$Be as a result of $^{7}$Be(p,$\gamma$) reaction rate variations, as listed in Tables 8 and 9, 
are rather large. Note, that the listed values correspond to a factor of 2 variation in the reaction rates.
However, the adopted $^{7}$Be(p,$\gamma$) reaction rate error amounts only to 12\% (Table 3),
yielding a $^{7}$Be abundance change of less than a factor of 2.
}
(Table 3). Only in model S1, which achieves the highest peak temperature, the $^{7}$Be abundance
changes by a factor of $\leq$20 as a result of $^{8}$B(p,$\gamma$) reaction rate variations.
However, we emphasize that $^{7}$Be is a very fragile nucleus which is easily destroyed at high stellar
temperatures. In this particular case, convection plays a crucial role as pointed out by
Hernanz et al. (1996). Consequently, the $^{7}$Be abundance could be far less sensitive to
$^{8}$B(p,$\gamma$) reaction rate variations in a hydrodynamic nova simulation. Such
studies are underway and the results will be reported in a forthcoming publication
(Starrfield et al. 2002).

We conclude that for nova models, with the possible exception of ONe nova model S1, estimates of 
Galactic $^{7}$Li production and of the 478 keV $\gamma$--ray line intensity from $^{7}$Be
decay are insensitive to present reaction rate uncertainties.

\subsubsection{Carbon Isotopes}

Models of ONe novae assume small initial $^{12}$C abundances, while the opposite is the case for
CO nova models (Table 2). Therefore, we expect the final carbon isotopic abundances in ONe and CO
nova models to depend on the rates of different reactions. This is indeed the case, as
can be seen from Tables 5--11. The $^{12}$C and $^{13}$C isotopic abundances show a dependence
on reaction rate variations of $^{13}$N(p,$\gamma$), $^{17}$O(p,$\gamma$), $^{17}$O(p,$\alpha$) and 
$^{17}$F(p,$\gamma$) in ONe nova models, and of $^{12}$C(p,$\gamma$), $^{13}$C(p,$\gamma$) and 
$^{14}$N(p,$\gamma$) in CO nova models. However, the abundances change by less than 50\% in all 
nova models considered in the present work. 

Present reaction rate estimates seem to be reliable for predicting carbon abundances
in nova ejecta and $^{12}$C/$^{13}$C isotopic abundance ratios of presolar grains originating from novae.

\subsubsection{Nitrogen Isotopes}

Abundances of the isotopes $^{14}$N and $^{15}$N show a dependence on reaction rate variations 
of $^{13}$N(p,$\gamma$), $^{14}$N(p,$\gamma$), $^{15}$N(p,$\alpha$), $^{17}$O(p,$\gamma$), 
$^{17}$O(p,$\alpha$), $^{17}$F(p,$\gamma$) and $^{18}$F(p,$\alpha$). The relative importance of these 
reactions depends
on the particular nova model considered. However, as was the case for carbon, changes in 
nitrogen abundances amount to less than 50\% in all models. 

Therefore, current reaction rates are sufficiently reliable for predictions of nitrogen
abundances in nova ejecta and of $^{14}$N/$^{15}$N abundance ratios of presolar grains originating from novae.

\subsubsection{Oxygen Isotopes}

For CO nova models, oxygen abundances show a weak dependence on variations in $^{16}$O(p,$\gamma$) 
and $^{17}$O(p,$\gamma$) reaction rates, with abundance changes of less than a factor of two. 
However, $^{17}$O abundances are sensitive to the $^{17}$O(p,$\alpha$) reaction rate.
Variations of the corresponding reaction rates give rise to $^{17}$O abundance changes by 
factors of $\leq$30.

In ONe nova models P1, P2 and S1, variations in $^{18}$F(p,$\alpha$) reaction rates change 
$^{16}$O abundances by factors of $\leq$50. 
Abundances of $^{17}$O are sensitive to reaction rate variations of $^{17}$F(p,$\gamma$)
in models JCH2, P1, P2, and S1, 
resulting in abundance changes by factors of $\leq$500.
In models JCH1, JCH2 and P1, the abundance of $^{17}$O changes by factors of $\leq$170 as a result
of varying the $^{17}$O(p,$\alpha$) reaction rates. The $^{17}$O abundance is also influenced
by rate variations of $^{17}$O(p,$\gamma$) in models JCH1 and JCH2, and of $^{18}$F(p,$\alpha$) in model S1,
resulting in abundance changes by factors of $\leq$6 and $\leq$15, respectively.
Note, that the final abundance of $^{18}$O originates
predominantly from the decay of $^{18}$F and, therefore, the abundance changes of both isotopes
will depend on the rates of the same reactions. The isotope $^{18}$F is discussed below.

Clearly, the rates of several reactions have to be improved in order to predict both more reliable
oxygen abundances in nova ejecta and $^{16}$O/$^{17}$O ratios of presolar grains originating from novae.

\subsubsection{Isotope $^{18}$F}

For CO nova models, $^{18}$F abundances are sensitive to $^{18}$F(p,$\alpha$), $^{17}$O(p,$\alpha$)
and $^{17}$O(p,$\gamma$) reaction rate variations, yielding abundance changes by factors 
of $\leq$100.

For all ONe nova models considered here, $^{18}$F abundances depend sensitively on variations in
$^{17}$O(p,$\gamma$) and $^{18}$F(p,$\alpha$) reaction rates, with
abundance changes by factors of $\leq$500. In models JCH1, JCH2 and P1, variations in 
$^{17}$O(p,$\alpha$) reaction rates change $^{18}$F abundances by factors of $\leq$110.
The $^{17}$F(p,$\gamma$) reaction also influences the $^{18}$F abundance in models
JCH2, P1, P2 and S1, resulting in abundance changes by factors of $\leq$600.

In summary, the $^{18}$F abundance is sensitive to present reaction rate uncertainties in all
nova models considered here. Consequently, the rates of several reactions have to be improved
in order to predict with more confidence the early $\gamma$--ray emission from novae at and below 511 keV.

\subsection{Mass Region A$\geq$20}

\subsubsection{Neon Isotopes}

Isotopic abundances of $^{20}$Ne and $^{21}$Ne depend only weakly on reaction rate variations
in CO nova models. The $^{22}$Ne abundance is sensitive to $^{22}$Ne(p,$\gamma$) reaction 
rate variations. Corresponding abundance changes amount to factors of $\leq$100. 

For ONe nova models, effects of reaction rate variations on the isotopic abundances of
$^{20}$Ne, $^{21}$Ne and $^{22}$Ne depend on the peak temperature achieved (Table 1). 
For model S1, which achieves the highest
peak temperature, the abundance of the isotope $^{20}$Ne is
sensitive to variations of $^{23}$Na(p,$\gamma$) and $^{23}$Mg(p,$\gamma$) reaction rates.
Abundance changes amount to factors of $\leq$11. 
Abundances of $^{21}$Ne are sensitive to reaction rate variations of $^{21}$Na(p,$\gamma$) in model
P2 and of $^{21}$Na(p,$\gamma$), $^{23}$Na(p,$\gamma$) and $^{23}$Mg(p,$\gamma$) in model S1, resulting in 
abundance changes by factors of $\leq$13. The $^{22}$Ne abundance
is sensitive to $^{22}$Ne(p,$\gamma$) reaction rate variations in models JCH1, JCH2 and P1, and to
$^{21}$Na(p,$\gamma$) reaction rate variations in model P2. These abundance changes amount
to several orders--of--magnitude.

The dominant neon isotope in nova ejecta is $^{20}$Ne. In most nova models, its abundance is
insensitive to present reaction rate uncertainties. Only ONe nova models which achieve very high peak 
temperatures (for example, model S1) require improved reaction rates for the prediction of
accurate $^{20}$Ne abundances. Calculations of $^{20}$Ne/$^{21}$Ne and $^{20}$Ne/$^{22}$Ne isotopic
ratios of presolar grains originating from ONe novae also require improved rates for several reactions.

\subsubsection{Sodium Isotopes}

The $^{22}$Na abundance predicted by CO nova models depends only weakly on reaction rate
variations. Furthermore, only small amounts of $^{22}$Na are produced as compared to ONe nova 
models which are discussed below. The $^{23}$Na abundance is sensitive to $^{22}$Ne(p,$\gamma$)
reaction rate variations. Corresponding abundance changes amount to factors of $\leq$7.

For ONe nova models JCH1, JCH2 and P1, reaction rate variations have only minor effects on $^{22}$Na 
abundances. In models P2 and S1, which achieve the highest peak temperatures, variations of 
$^{21}$Na(p,$\gamma$) reaction rates have the effect of changing $^{22}$Na abundances by factors of 
$\approx$6. Variations in $^{23}$Na(p,$\gamma$) and $^{23}$Mg(p,$\gamma$) reaction rates have also
an effect in model S1, changing $^{22}$Na abundances by factors of $\leq$10. The $^{23}$Na
abundance is sensitive to $^{23}$Na(p,$\gamma$) reaction rate variations in all ONe nova models,
resulting in abundance changes by factors of $\leq$6. In models P2 and S1, the $^{23}$Na
abundance changes by factors of $\leq$7 if the $^{23}$Mg(p,$\gamma$) reaction rates are varied
within their errors.

For ONe nova models which achieve high peak temperatures, improved rates for several reactions are desirable
for estimating the intensity of the $\gamma$--ray line at 1275 keV originating from the decay
of $^{22}$Na. Furthermore, present reaction rate uncertainties have to be reduced 
in order to calculate reliable $^{23}$Na abundances in CO and ONe nova ejecta.

\subsubsection{Magnesium Isotopes}

For CO nova models, variations of $^{22}$Ne(p,$\gamma$) and $^{23}$Na(p,$\gamma$) reaction rates
change $^{24}$Mg abundances by factors of $\leq$70. The $^{25}$Mg abundance depends on the
$^{22}$Ne(p,$\gamma$) reaction rate in model JH2, resulting in abundance changes by factors
of $\leq$5. Variation of the rates for $^{26}$Mg(p,$\gamma$)
changes the abundance of $^{26}$Mg by factors of $\leq$14. 

For all ONe nova models considered here, variations of $^{23}$Na(p,$\gamma$) reaction rates change
the abundances of $^{24}$Mg, $^{25}$Mg and $^{26}$Mg by factors of $\leq$60. 
In models P2 and S1, which achieve the highest peak temperatures, the $^{24}$Mg abundance
depends also on $^{23}$Mg(p,$\gamma$) reaction rate variations, resulting in abundance changes by factors
of $\leq$7. In all ONe nova models, the $^{26}$Mg abundance changes by factors of $\leq$14 as a 
result of $^{26}$Al$^{m}$(p,$\gamma$) reaction rate variations.
In models JCH1 and JCH2, the $^{26}$Mg abundance is also sensitive to $^{26}$Mg(p,$\gamma$) rate
variations. Corresponding abundance changes amount to factors of $\leq$8. Note, that all
reaction rate variations tend to decrease magnesium isotopic abundances.

In summary, several reaction rate uncertainties have to be reduced in order to calculate accurate
magnesium isotopic abundances.

\subsubsection{Aluminum Isotopes}

Variations of $^{22}$Ne(p,$\gamma$) and $^{26}$Al$^{g}$(p,$\gamma$) reaction rates change
$^{26}$Al abundances by factors of $\leq$20 and $\leq$5, respectively. However, CO nova models
produce smaller amounts of $^{26}$Al compared to ONe models which are discussed
below. The abundance of $^{27}$Al depends only weakly on reaction rate variations in CO nova models.

For all ONe nova models, $^{26}$Al abundances are sensitive to $^{23}$Na(p,$\gamma$) and 
$^{26}$Al$^{g}$(p,$\gamma$) reaction rate variations, yielding abundance changes by factors of $\leq$60.
Variations in $^{23}$Na(p,$\gamma$) reaction rates change $^{27}$Al abundances by factors
of $\leq$60.

Clearly, certain reaction rates have to be improved in order to predict not only more reliable
aluminum abundances in nova ejecta, but also to estimate the contribution of novae to the Galactic $^{26}$Al abundance
and $^{26}$Al/$^{27}$Al ratios of presolar grains originating from novae.

\subsubsection{Silicon Isotopes}

The nucleosynthesis of silicon isotopes in CO nova models is negligible and reaction rate
variations have only insignificant effects.

For all ONe nova models considered, the $^{28}$Si abundance is insensitive to reaction rate variations.
All models predict a dependence of $^{29}$Si abundances on $^{29}$Si(p,$\gamma$)
reaction rate variations, with abundance changes by factors of $\leq$14. In all models, variations in 
$^{30}$P(p,$\gamma$) reaction rates have the effect of changing $^{30}$Si abundances by factors of $\leq$100.

In conclusion, improved rates for several reactions are desirable in ONe nova models in order to estimate 
accurate silicon abundances in nova ejecta and silicon isotopic ratios of presolar grains originating from novae.

\subsubsection{Sulfur Isotopes}

Similar to the case of silicon, the nucleosynthesis of sulfur isotopes in CO nova
models is negligible.

For ONe nova models JCH1, JCH2, P1 and P2, reaction rate variations of $^{30}$P(p,$\gamma$) have the
effect of changing $^{32}$S abundances by factors of $\leq$12. 
For all models, variations in $^{33}$S(p,$\gamma$) reaction rates change $^{33}$S abundances 
by factors of $\leq$1000. The $^{33}$S abundance also depends on reaction rate variations of
$^{30}$P(p,$\gamma$) in models JCH1, JCH2, P1 and P2, and of $^{33}$Cl(p,$\gamma$) in models P2 and S1.
Abundance changes amount to factors of $\leq$14. The $^{34}$S abundance depends in all models
on $^{34}$S(p,$\gamma$) reaction rate variations, resulting in abundance changes by factors
of $\leq$130. The $^{34}$S abundance also depends on reaction rate variations of
$^{30}$P(p,$\gamma$) in models JCH2, P1 and P2, of $^{33}$S(p,$\gamma$) in models JCH2, P1, P2 and S1,
and of $^{34}$Cl(p,$\gamma$) in models P2 and S1. Abundance changes amount to factors of $\leq$13,
$\leq$30 and $\leq$5, respectively.

Consequently, uncertainties of several reaction rates have to be reduced in ONe nova models for the prediction
of accurate sulfur abundances in nova ejecta and of sulfur isotopic ratios of presolar grains
originating from novae.

\subsubsection{Chlorine Isotopes}

Variations in $^{34}$S(p,$\gamma$) reaction rates increase $^{35}$Cl abundances by factors
of $\leq$5 in CO nova models. The nucleosynthesis of $^{37}$Cl is negligible and reaction rate variations
have only insignificant effects. 
Note, that CO nova models produce much less $^{35}$Cl compared to ONe nova models
which are discussed below.

In ONe nova models JCH2, P1 and P2, the $^{35}$Cl abundance changes by factors of $\leq$10 as
a result of varying the $^{30}$P(p,$\gamma$) and $^{33}$S(p,$\gamma$) reaction rates.
The $^{35}$Cl abundance is also sensitive to $^{34}$S(p,$\gamma$) reaction rate variations
in models JCH1, JCH2, P1 and P2, resulting in abundance changes by factors of $\leq$20.
The abundance of $^{37}$Cl changes in model P2 by factors of $\leq$24 as a 
result of $^{30}$P(p,$\gamma$) and $^{37}$Ar(p,$\gamma$) reaction rate variations, while in
model S1 abundance changes of $^{37}$Cl amount to factors of $\leq$38 as a result of varying
the $^{37}$Ar(p,$\gamma$) reaction rates. Most reaction rate variations tend to decrease
the abundances of $^{35}$Cl and $^{37}$Cl.

Therefore, the calculation of reliable chlorine abundances in the ejecta of ONe novae requires improved
rates for several reactions.

\subsubsection{Argon Isotopes}

The nucleosynthesis of argon isotopes in CO nova models is negligible and reaction rate
variations have only insignificant effects.

For ONe nova models P1 and P2, the $^{36}$Ar abundance changes by factors of $\leq$7 if the
$^{30}$P(p,$\gamma$), $^{33}$S(p,$\gamma$) and $^{34}$S(p,$\gamma$) reaction rates are
varied within their errors. The $^{37}$Ar abundance is sensitive to $^{37}$Ar(p,$\gamma$)
reaction rate variations in all ONe nova models, and to $^{30}$P(p,$\gamma$) rate
variations in model P2. Abundance changes amount to factors of $\leq$120. 
The $^{38}$Ar abundance is sensitive to $^{37}$Ar(p,$\gamma$) reaction rate variations 
in all models, to $^{38}$K(p,$\gamma$) reaction rate variations in models JCH2, P1, P2 and S1,
and to variations of $^{30}$P(p,$\gamma$) reaction rates in model P2. 
Changes in $^{38}$Ar abundances amount to factors of $\leq$18. 
As was the case for chlorine isotopes, most reaction rate variations tend to decrease the 
argon isotopic abundances.

In conclusion, several reaction rate uncertainties have to be reduced in order to predict
reliable argon abundances in ONe nova ejecta.

\subsection{Comparison with Hydrodynamic Model Calculations}

In $\S$ 2 we have pointed out that hydrodynamic nova model calculations are time consuming and,
consequently, the effect of reaction rate variations on final isotopic abundances has been
previously studied for only a few selected cases. In the following, we will compare some of our results
with those obtained from previous hydrodynamic model calculations. It has to be kept in mind, as
already discussed in detail, that our calculations neglect convection. 

The dependence of $^{18}$F abundances on reaction rate variations has been studied by 
Coc et al. (2000). Their hydrodynamic model calculations were performed for ONe nova model
JCH2 which has also been used for the one--zone calculations of the present work (Table 1).
They quote factors of 10 and 310 for the ratio of maximum versus minimum $^{18}$F 
abundance as a consequence of $^{17}$O+p and $^{18}$F+p reaction rate variations, respectively.
Our result for $^{18}$F+p reaction rate variations is in agreement with that of Coc et al. (2000),
but for variations of $^{17}$O+p reaction rates we obtain larger $^{18}$F abundance changes.
The agreement for $^{18}$F+p and the disagreement for $^{17}$O+p could be explained by the
fact that we use the same $^{18}$F+p reaction rates and corresponding errors as Coc et al.
(2000), while for $^{17}$O+p we use newer reaction rates which differ significantly
from those adopted previously ($\S\S$ 3.1 and 3.3).

Abundance changes of $^{22}$Na as a result of reaction rate variations have been studied by 
Jos\'{e} et al. (1999).
They performed hydrodynamic ONe nova model calculations assuming white dwarf masses of
1.15 M$_{\odot}$ and 1.25 M$_{\odot}$. The models are described in detail in 
Jos\'{e} \& Hernanz (1998). They quote an increase in $^{22}$Na abundance by a factor of 2--3
as a result of reducing the $^{21}$Na(p,$\gamma$) reaction rates (adopted from
Caughlan \& Fowler 1988) by a factor of 100. For ONe nova models JCH1 and JCH2, which are similar
to those used in Jos\'{e} et al. (1999), we 
also observe an increase of $^{22}$Na abundance as a result of $^{21}$Na(p,$\gamma$) 
reaction rate decreases, although we find smaller effects in our one--zone calculations 
($\leq$50\% decrease of abundance). The difference might be explained by the fact that some 
of the key reaction rates in this mass range adopted in Jos\'{e} et al. (1999) and in the 
present work differ significantly.

Finally, the effects of $^{30}$P(p,$\gamma$) reaction rate variations on the synthesis
of elements between Si and Ca has been investigated in Jos\'{e} et al. (2001).
They adopted a hydrodynamic ONe nova model with a white dwarf mass of 1.35 M$_{\odot}$,
reaching a peak temperature of T$_{peak}$=0.331 GK. Details of the model can be found
in Jos\'{e} \& Hernanz (1998).
They quote that the abundances of several isotopes ($^{31}$P, $^{32}$S, $^{33}$S, $^{34}$S,
$^{35}$Cl and $^{36}$Ar) decrease by about an order of magnitude when the
$^{30}$P(p,$\gamma$) reaction rates are decreased by a factor of 100. They also find that
only the $^{30}$Si abundance changes by significant amounts if the
$^{30}$P(p,$\gamma$) reaction rates are increased by a factor of 100 (Table 2 in 
Jos\'{e} et al. 2001). Note, that the ONe nova models considered in the present work (Table 1) 
are different 
from the one adopted in Jos\'{e} et al. (2001). Nevertheless, for $^{30}$P(p,$\gamma$) reaction rate
variations we find qualitative and quantitative agreement with Jos\'{e} et al. (2001)
for all ONe nova models, as can be seen from Tables 5--11 (see also Fig. 3d).

\section{SUMMARY AND CONCLUSIONS}

In the present work, we have investigated the effects of thermonuclear reaction rate uncertainties on nova 
nucleosynthesis. One--zone nucleosynthesis calculations have been performed by adopting
temperature--density--time profiles of the hottest hydrogen--burning zone 
from 7 different, recent hydrodynamic nova simulations (Politano et al. 1995; 
Jos\'{e} \& Hernanz 1998; Jos\'{e}, Coc, \& Hernanz 1999;
Starrfield et al. 2002). The adopted nova models cover peak
temperatures in the range of T$_{peak}$=0.145--0.418 GK (Table 1). For each of these 
temperature--density--time profiles we have individually 
varied the rates of 175 reactions within their associated errors (Table 3) and 
analyzed the resulting abundance changes of 142 isotopes in the mass range below A=40.
In total, we performed $\approx$7350 reaction network calculations.
We use the most recent thermonuclear reaction rate
evaluations for the mass ranges A=1--20 (Angulo et al. 1999) and A=20--40 (Iliadis et al. 2001).
Results are presented in tabular form for each adopted nova simulation (Tables 5--11).
Figure 3 displays the results of reaction rate variations for a few selected cases.
We find that present reaction rate estimates are reliable for predictions of
Li, Be, C and N abundances in nova nucleosynthesis. However, uncertainties in the rates of several reactions
have to be reduced significantly in order to predict more reliable O, F, Ne, Na, Mg, Al, Si, S,
Cl and Ar abundances. 

It is important to emphasize how to interpret the results of the present work. 
Hydrodynamic nova model
calculations clearly show that typically only the outer layers of the envelope, 
not the deepest layers of the hydrogen--burning shell, are ejected after the thermonuclear runaway. 
The ejected layers are enriched, through convective mixing, with the products of 
the inner hydrogen--burning shell. From these considerations, it is clear that our
calculations are unsuitable for defining {\it absolute} isotopic abundances resulting from
nova nucleosynthesis, since our one--zone calculations necessarily ignore
convection ($\S$ 2). Nevertheless, our procedure is adequate for exploring the effects of
reaction rate uncertainties on abundance {\it changes} in the hottest hydrogen--burning zone, i.e., 
the region in which most of the nucleosythesis takes place. It follows, therefore, that our final abundances
(Table 4) should neither be compared to elemental abundances observed in nova ejecta nor to
results from hydrodynamic model calculations. We also would like to stress the following point. 
If a particular reaction rate variation has insignificant effects on isotopic abundances in our calculations, 
then it is most likely that a full hydrodynamic model calculation will yield a similar result.
However, the reverse statement is not neccessarily correct, i.e., if we find significant
abundance changes as a result of a particular reaction rate variation, then a full hydrodynamic
model calculation might not produce significant effects. 
Clearly, our work does not represent the final answer to the question of which reactions should 
be targets for future measurements, but should be regarded as a first step in that direction.

In Table 12 we summarize qualitatively some of our results. The table lists isotopes
whose abundances change by more than a factor of 2 in at 
least one of the nova models considered here
as a result of varying a particular reaction rate within uncertainties.  
It is striking that for the vast majority of reactions included in our 
network calculations, reaction rate variations have an insignificant effect on final isotopic abundances 
in all nova models. Instead, final abundances are influenced by variations of a restricted number 
of key reaction rates. 
Closer inspection of Tables 5--11 also shows that variations of the same reaction rates
in nova models of the same white dwarf mass (e.g., models P1 and JCH2 with M$_{WD}$=1.25M$_{\odot}$;
or models P2 and S1 with M$_{WD}$=1.35M$_{\odot}$) yield quantitatively different 
changes in final abundances. This is not surprising since different nova models 
assume different initial envelope compositions (Table 2) and
achieve different peak temperatures (Table 1).

It can be seen from Table 12 and from Figure 3 that reaction rate variations of a few reactions, such as 
$^{23}$Na(p,$\gamma$)$^{24}$Mg, $^{23}$Mg(p,$\gamma$)$^{24}$Al, $^{30}$P(p,$\gamma$)$^{31}$S 
and $^{33}$S(p,$\gamma$)$^{34}$Cl, influence final abundances of a large number of
isotopes. 
Consequently, new measurements of these reactions could significantly reduce
uncertainties of isotopic abundances in nova model calculations. 
The reader might be surprised by the fact that certain reactions that were previously thought
to play a role in nova nucleosynthesis do not appear in Table 12. 
In agreement with previous work (Iliadis et al. 1999), 
we find insignificant isotopic abundance changes as a result
of $^{27}$Si(p,$\gamma$)$^{28}$P, $^{31}$S(p,$\gamma$)$^{33}$Cl, $^{35}$Ar(p,$\gamma$)$^{36}$K
and $^{39}$Ca(p,$\gamma$)$^{40}$Sc reaction rate variations for all nova models. This result has been
confirmed by recent hydrodynamic model calculations (Jos\'{e} et al. 2001). 
The $^{15}$O($\alpha$,$\gamma$)$^{19}$Ne and $^{19}$Ne(p,$\gamma$)$^{20}$Na reactions,
which were thought to cause a breakout of material from the CNO mass region to the region
beyond Ne, are also missing in Table 12. Rate variations for both reactions have only
small effects on final abundances in all nova models, except in model S1 which achieves the highest
peak temperature (T$_{peak}$=0.418 GK). According to Table 7,
an increase of those two reaction rates by a factor of 100 has only a moderate influence on abundance
changes in the mass range below A=20. But even for this rather high peak temperature, 
no breakout of material from the CNO mass region is observed. This result has also been
confirmed by recent hydrodynamic model calculations (Starrfield et al. 2002).
It is also apparent from Tables 5--11 that ($\alpha$,$\gamma$) and ($\alpha$,p)
reactions in general are not important for nova nucleosynthesis.

Finally, it should be noted that it is difficult to estimate
reliable reaction rate errors in certain cases. Consider as an example the $^{25}$Al(p,$\gamma$)$^{26}$Si reaction. 
In this case, as for most other reactions involving short--lived target nuclei, we have assumed
a reaction rate error of a factor of 100 up and down ($\S\S$ 3.3 and Table 3). 
An inspection of Tables 5--11 reveals only small abundance changes (within a factor of 2) as a result of varying
the corresponding reaction rates within a factor of 100.
However, for this particular case
we have only limited experimental information regarding the energies of unobserved low--energy 
resonances (Iliadis et al. 1996). Depending on the location of these resonances, the $^{25}$Al(p,$\gamma$)$^{26}$Si
reaction rates could increase by much more than 2 orders of magnitude. As a consequence, the
$^{26}$Al abundance will decrease significantly in all ONe nova models. Although not listed in Table 12,
it is clear from this discussion that measurements of reactions such as $^{25}$Al(p,$\gamma$)$^{26}$Si
are also desirable in order to improve predictions 
of nova nucleosynthesis.

\vspace{10mm}

The authors would like to thank A. Coc, M. Hernanz, R. Hix and M. Smith for stimulating
discussions. We are also grateful for the detailed review of this work by the referee, S. Shore.
This work was supported in part by the U. S. Department of Energy under Grant No. DE--FG02--97ER41041,
by CICYT--PNIE ESP98--1348 and DGES PB98--1183--C03--02, and by Grants from NASA and NSF to ASU.

\clearpage

\begin{figure}
\caption{
Temperature--density profiles for the hottest hydrogen--burning zone of CO novae (dashed lines) and ONe novae 
(solid lines). 
The nuclear--burning conditions evolve in time from larger to smaller densities.
The profiles have been adopted from recently published hydrodynamic nova simulations and are described in
more detail in Table 1. The very small ripples visible in some profiles near peak temperature
originate from numerical instabilities which are not important for present considerations.
}
\end{figure}

\begin{figure}
\caption{
Initial envelope composition (in mass fractions) for nova nucleosynthesis calculations. 
(a) ONe models P1, P2, S1. (b) ONe models JCH1, JCH2. (c) CO model JH1. (d) CO model JH2. 
For more details, see Table 2. 
}
\end{figure}

\begin{figure}
\caption{
Factor change, X$_{i}$/X$_{i,rec}$, in final isotopic abundance as a result of varying a specific
reaction rate within the assigned errors (as given in Table 3) vs. mass number. The diamonds 
and triangles correspond to the upper and lower limit of the reaction rate, respectively.
(a) Variation of $^{21}$Na(p,$\gamma$)$^{22}$Mg reaction rate in ONe model P2.
(b) Variation of $^{23}$Na(p,$\gamma$)$^{24}$Mg reaction rate in ONe model JCH1.
(c) Variation of $^{23}$Mg(p,$\gamma$)$^{24}$Al reaction rate in ONe model S1.
(d) Variation of $^{30}$P(p,$\gamma$)$^{31}$S reaction rate in ONe model JCH2.
(e) Variation of $^{33}$S(p,$\gamma$)$^{34}$Cl reaction rate in ONe model P1.
The symbols for $^{22}$Na, $^{26}$Al and $^{37}$Ar have been shifted slightly to the right 
for clarity.
}
\end{figure}

\clearpage

\begin{deluxetable}{lccccccc}
\tiny
\tablecaption{PROPERTIES OF RECENT EVOLUTIONARY NOVA MODELS \tablenotemark{a}} 
\tablehead{ \colhead{MODEL} & \colhead{P1} & \colhead{P2} & \colhead{S1} & 
            \colhead{JCH1} & \colhead{JCH2} & \colhead{JH1}   & \colhead{JH2} \nl
            \colhead{WD MASS (M$_{\odot}$)} & \colhead{1.25} & \colhead{1.35} & \colhead{1.35} &
            \colhead{1.15} & \colhead{1.25} & \colhead{0.8} & \colhead{1.0} \nl
            \colhead{WD COMPOSITION} & \colhead{ONe} & \colhead{ONe} & \colhead{ONe} & \colhead{ONe} &
            \colhead{ONe} & \colhead{CO} & \colhead{CO} \nl
            \colhead{MIXING (\%)\tablenotemark{b}} &  \colhead{50} & \colhead{50} & \colhead{50} & \colhead{50} &
            \colhead{50} & \colhead{25} & \colhead{50}
} 
\startdata
T$_{peak}$ (10$^{6}$ K)                  & 290  & 356  & 418  & 231  & 251  & 145  & 170 \nl
L$_{peak}$ (10$^{4}$L$_{\odot}$)         & 4.3  & 16.3 & 39   & 26   & 46   & 3.5  & 23  \nl
M$_{acc}$ (10$^{-5}$M$_{\odot}$)         & 3.2  & 1.5  & 3.8  & 3.2  & 2.2  & 9.7  & 3.9 \nl
\.{M$_{acc}$} (10$^{-10}$M$_{\odot}$/yr) & 16   & 16   & 1.6  & 2.0  & 2.0  & 2.0  & 2.0 \nl
M$_{ej}$ (10$^{-5}$M$_{\odot}$)          & 0.0  & 0.62 & 2.2  & 2.6  & 1.8  & 7.0  & 2.3 \nl
\enddata
\tablenotetext{a}{Models labeled ``P", ``S", ``JCH" and ``JH" are adopted from Politano et al. (1995),
Starrfield et al. (2002),
Jos\'{e}, Coc, \& Hernanz (1999) and Jos\'{e} \& Hernanz (1998), respectively.}
\tablenotetext{b} {Percentage of mixing assumed between solar accreted matter with white dwarf core material. The initial
envelope composition is given in Table 2.}
\end{deluxetable}%

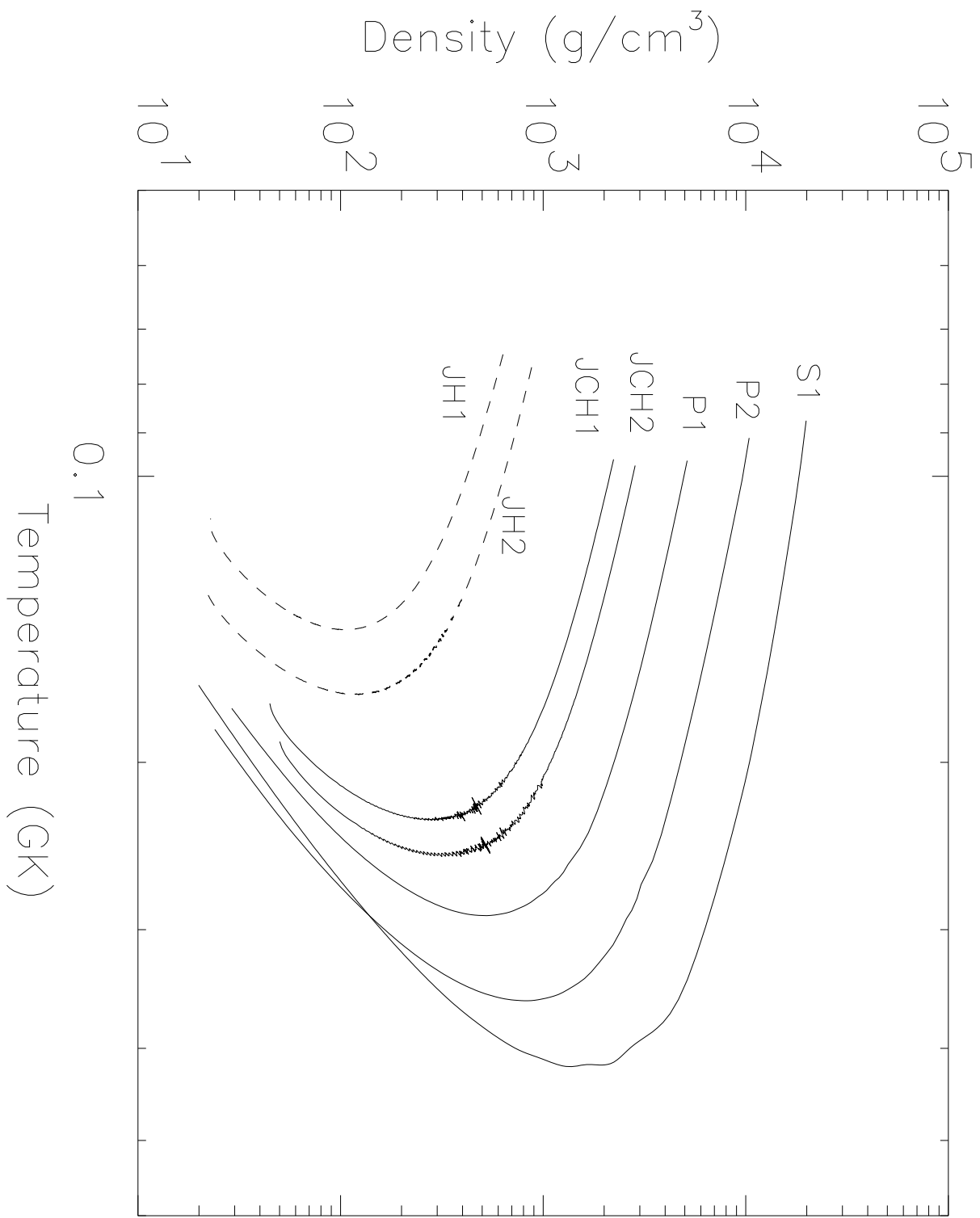
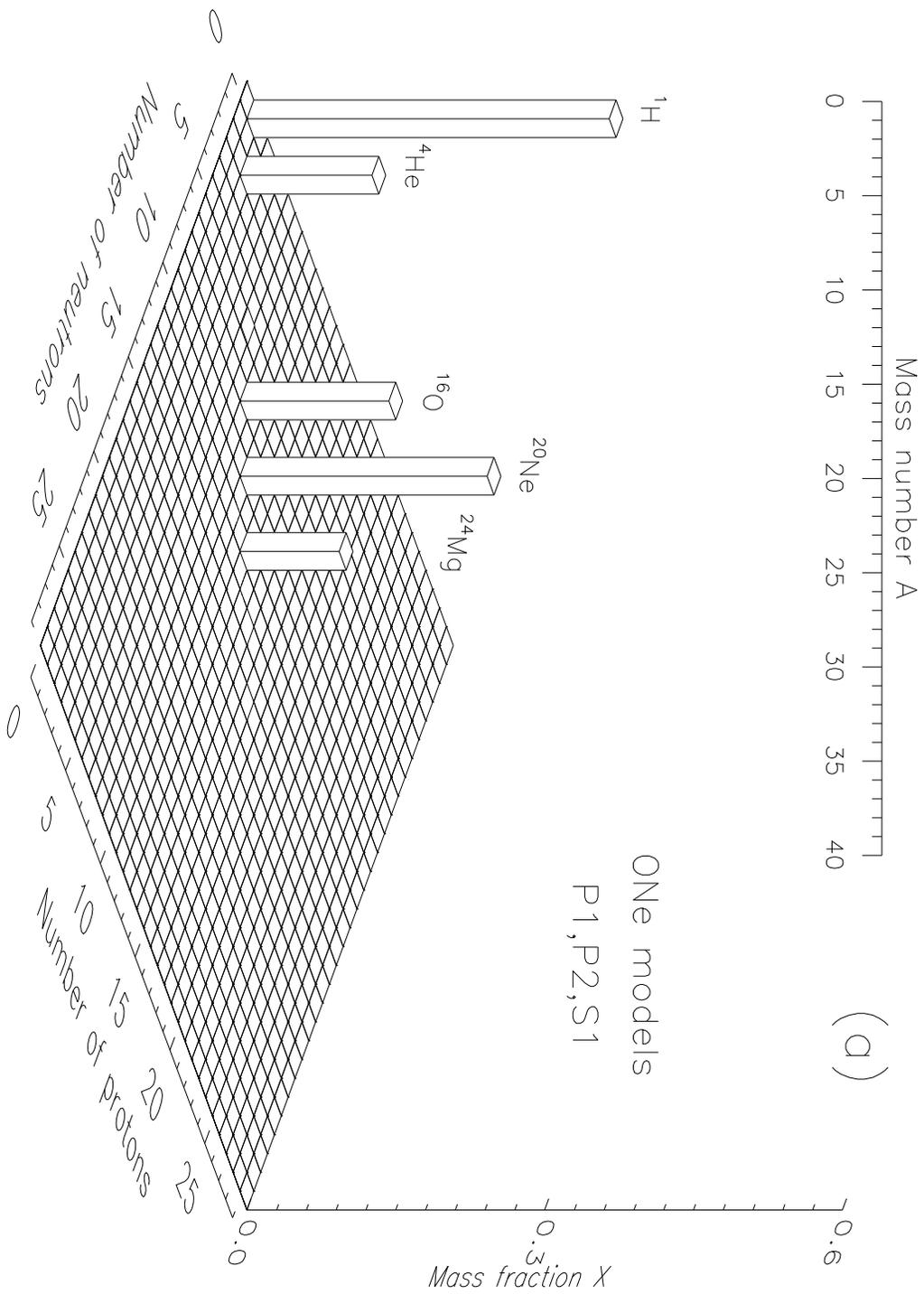
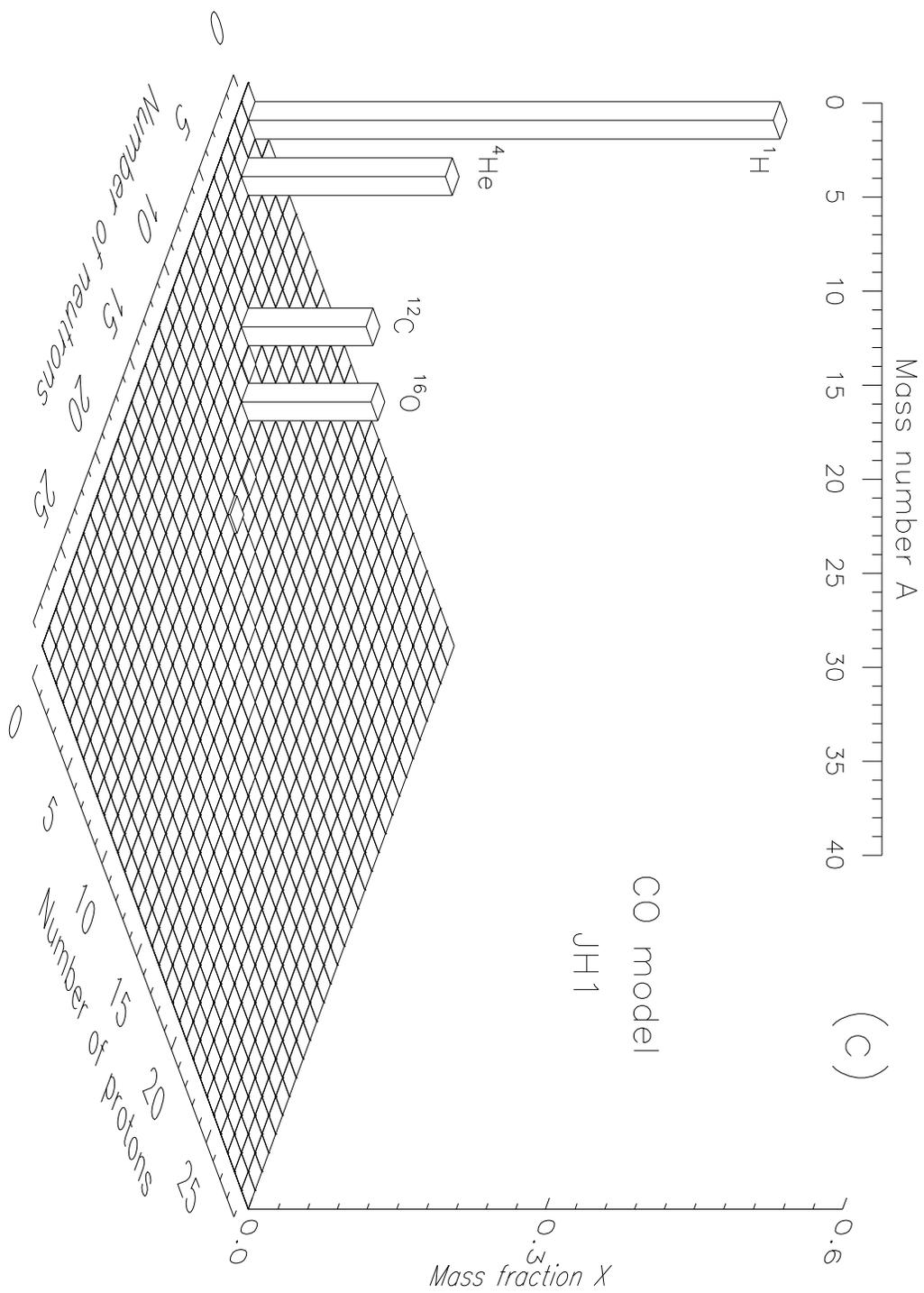
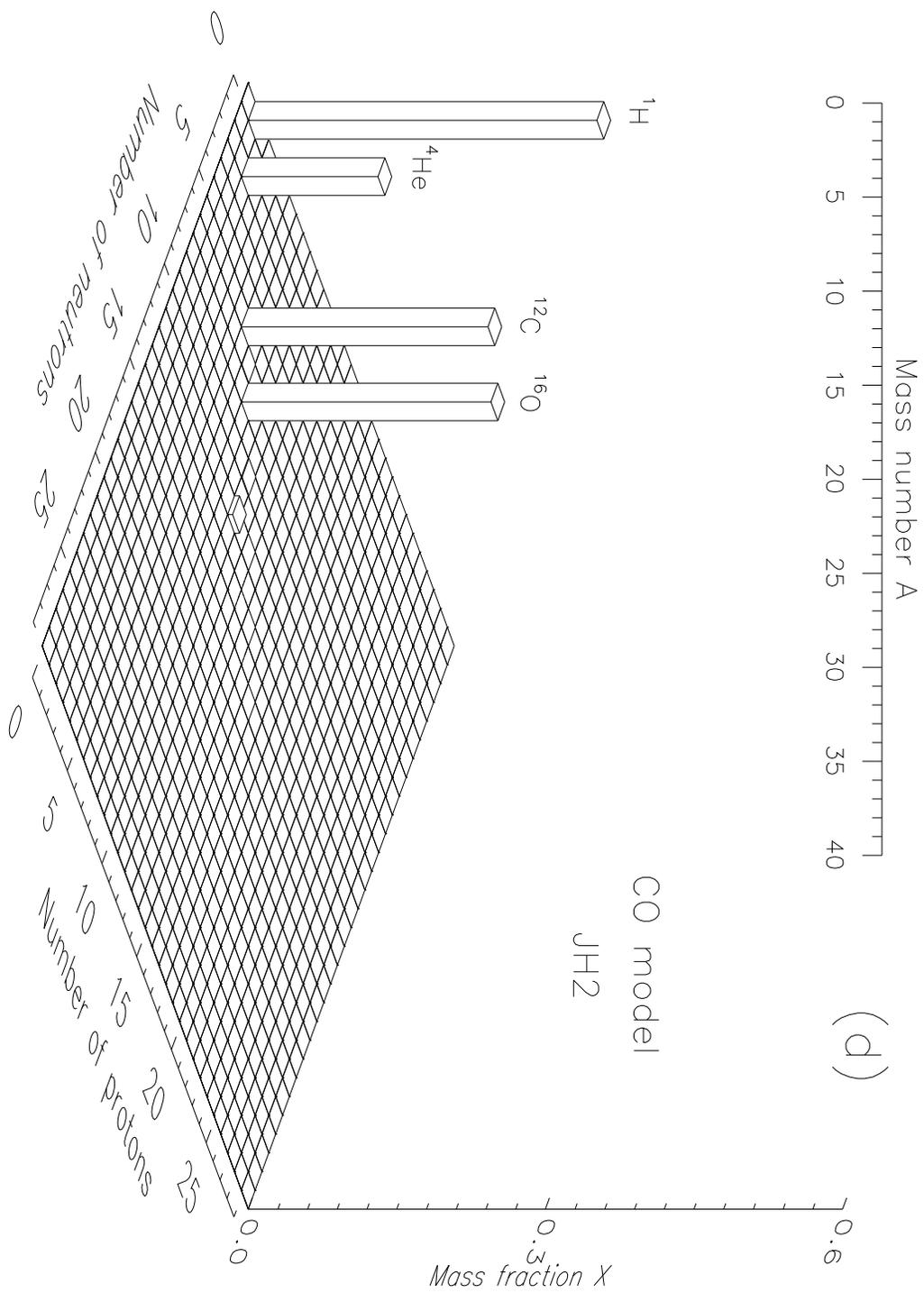
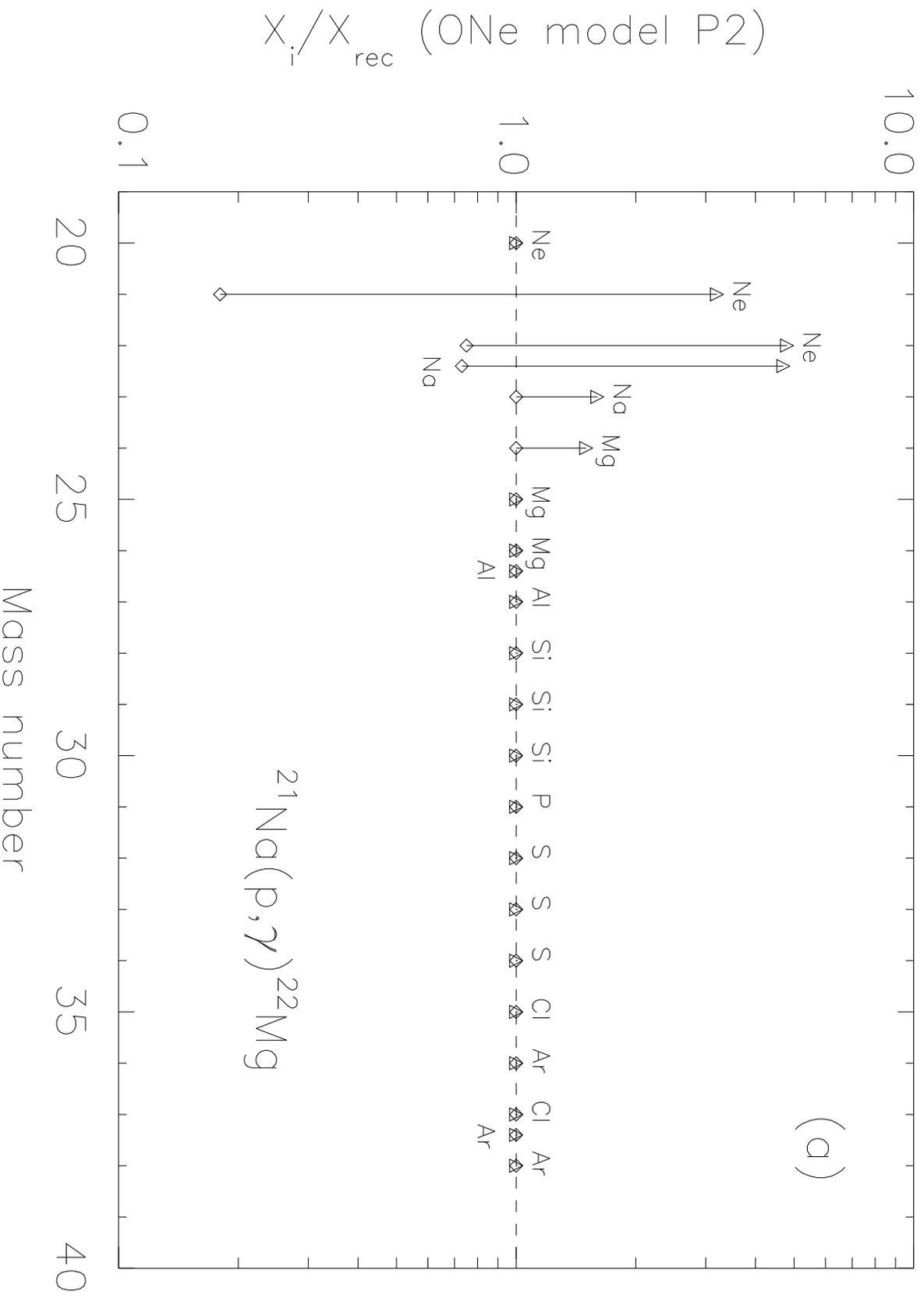
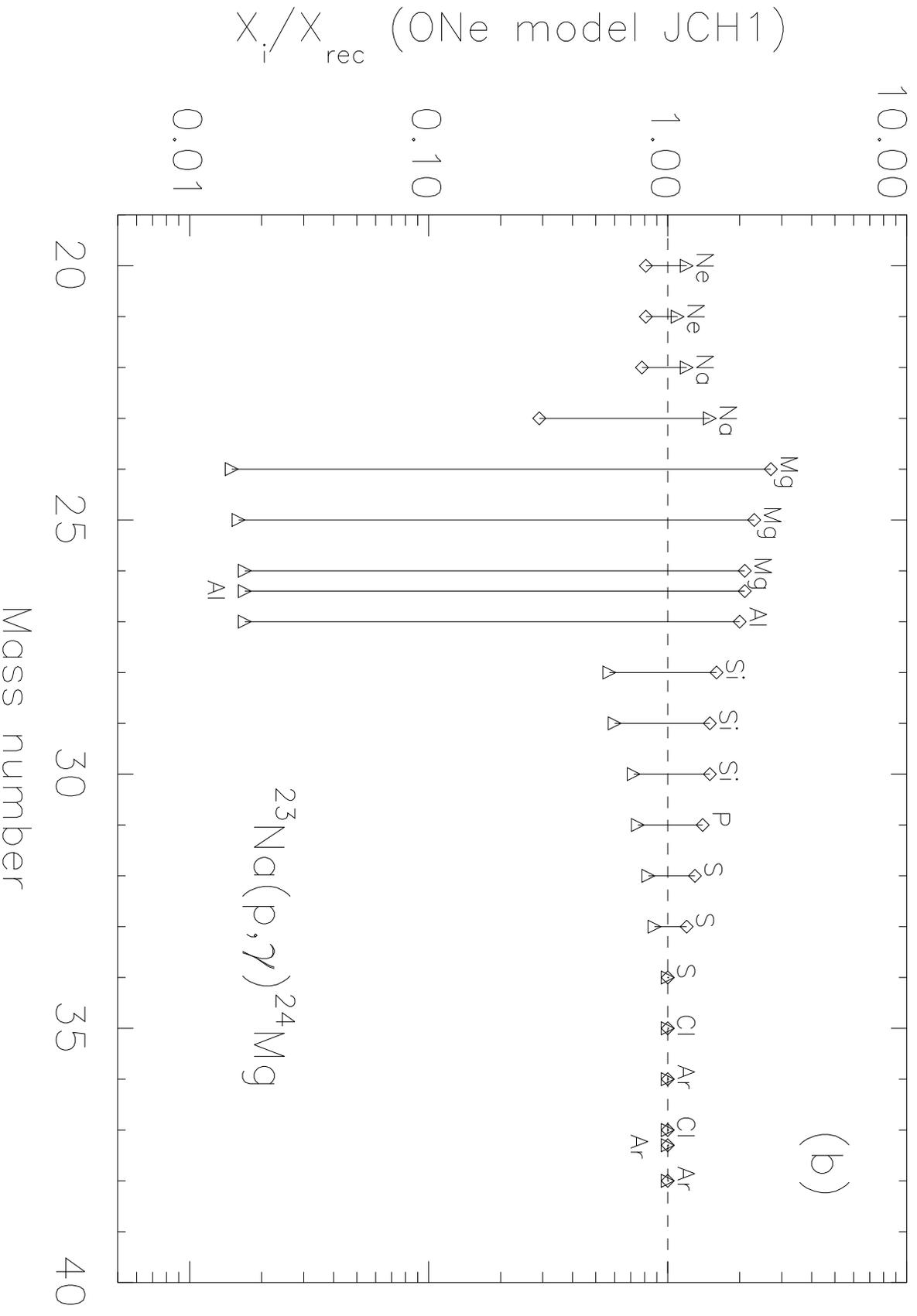
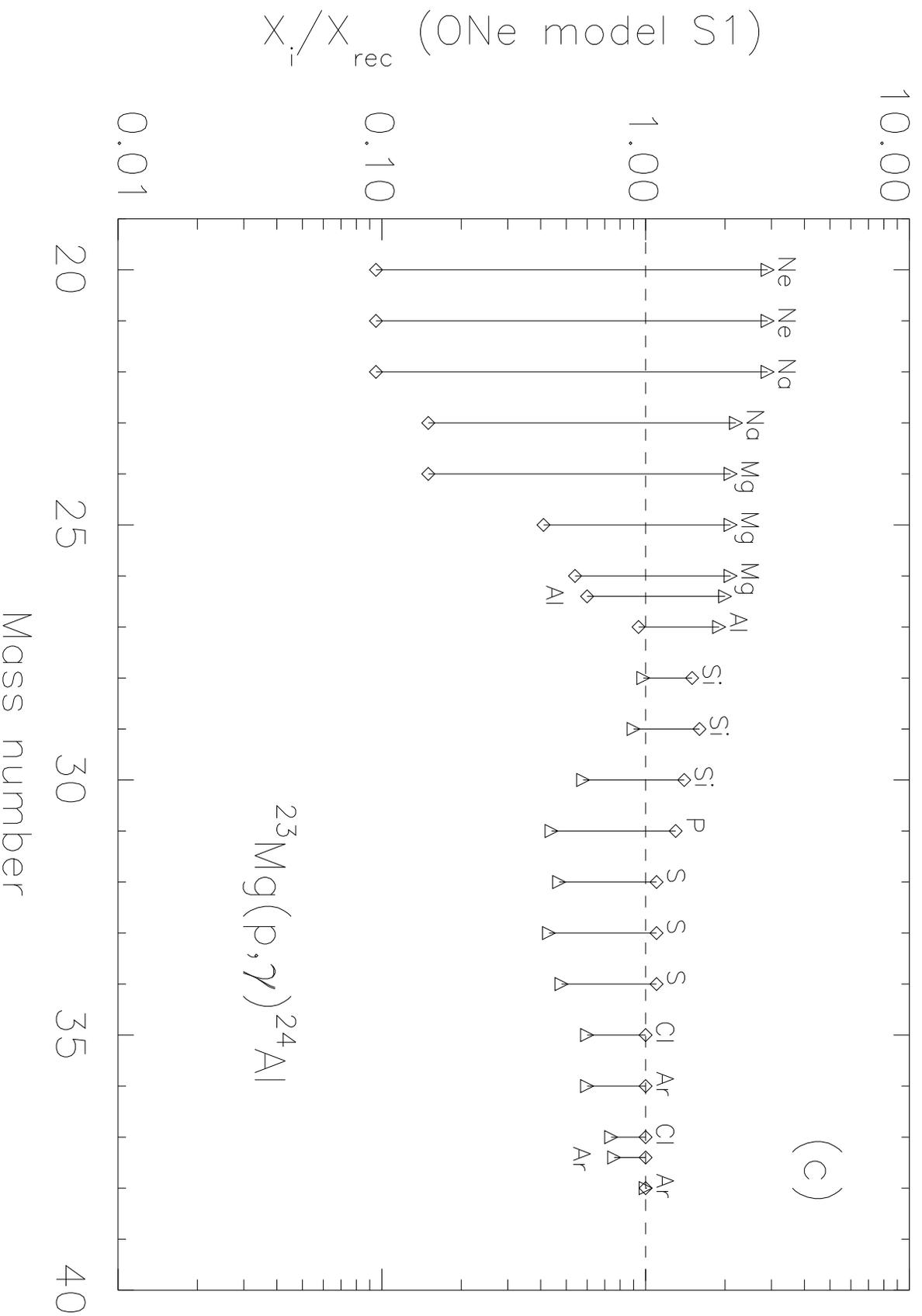
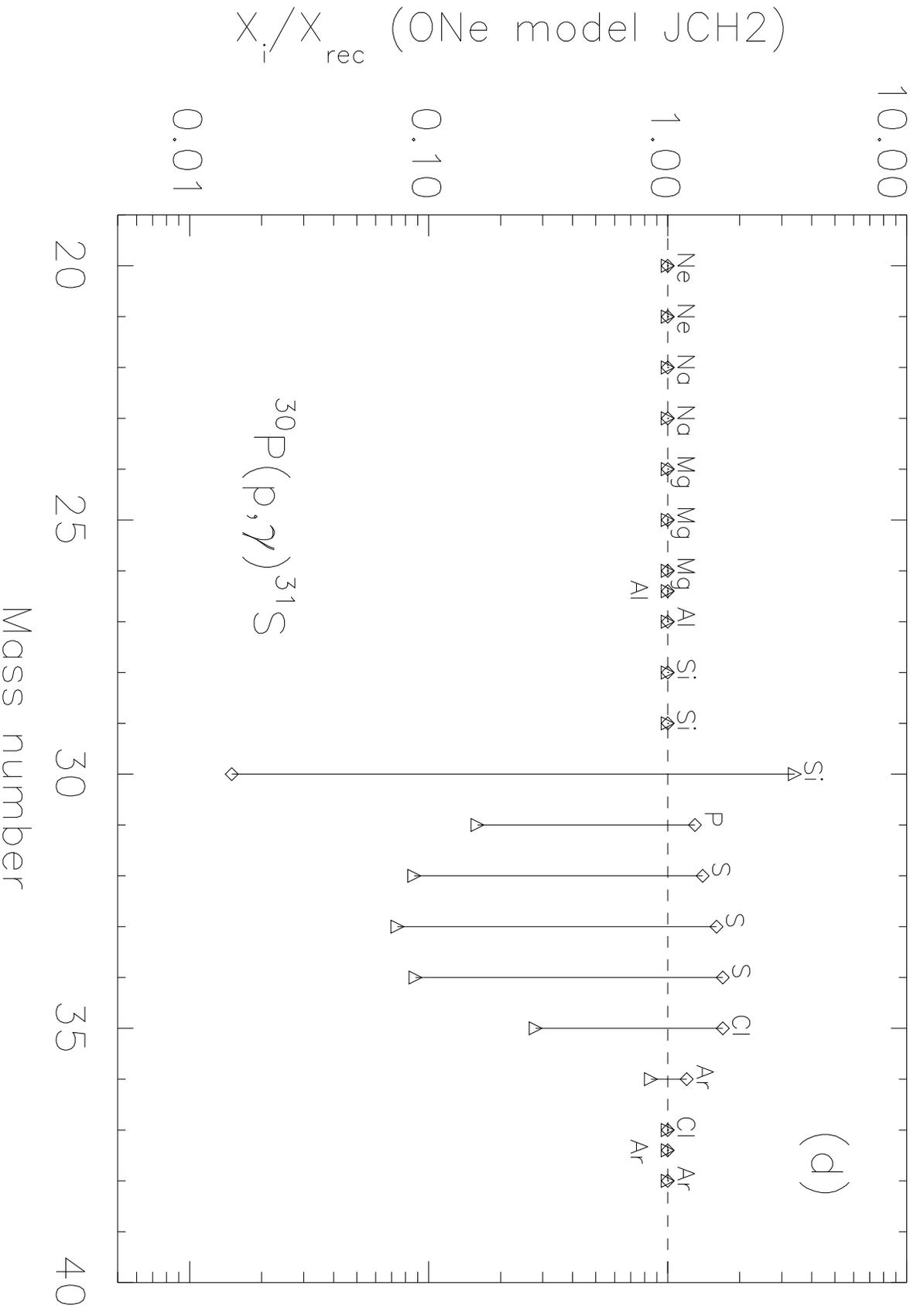
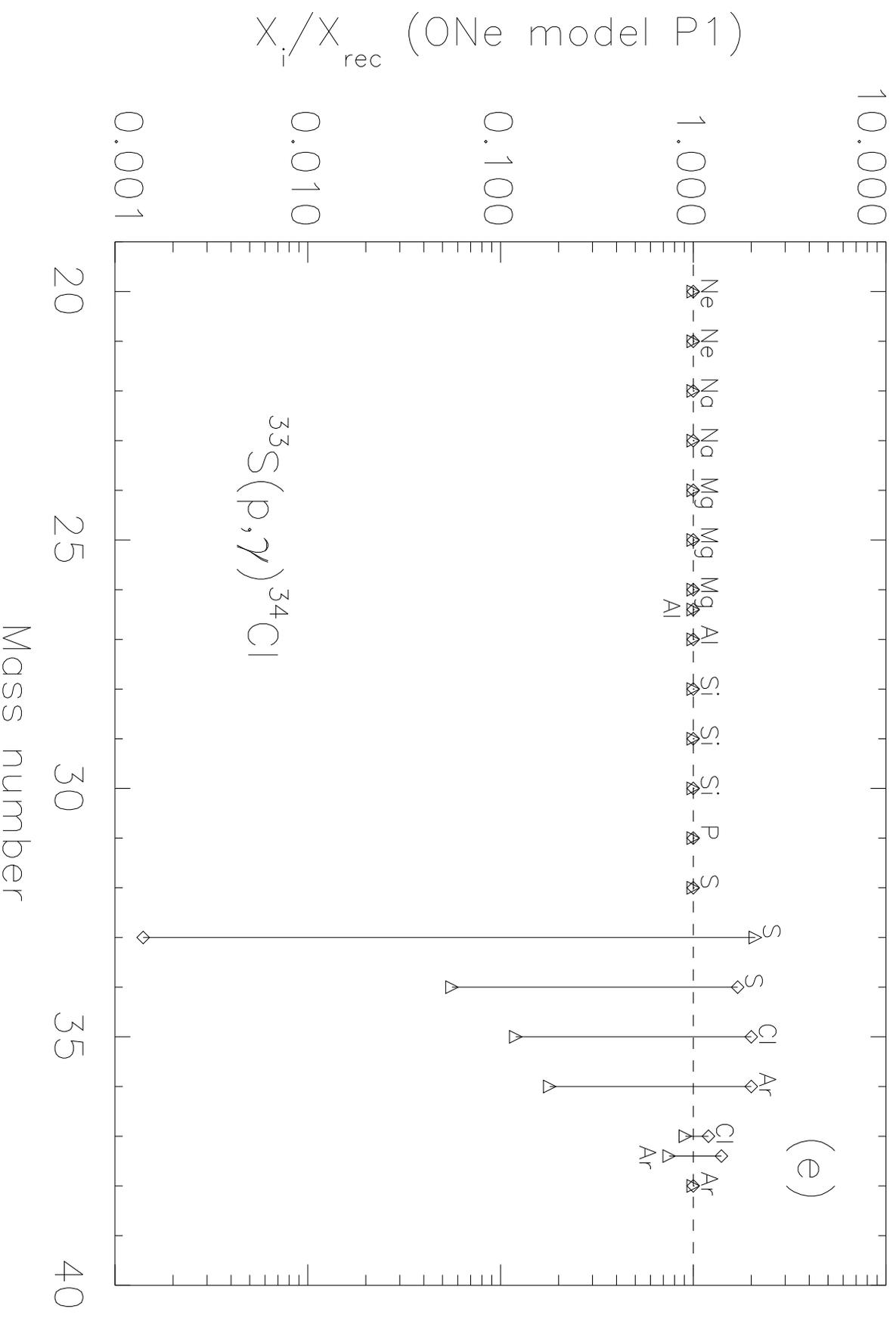
\end{document}